\documentclass[twocolumn,showkeys,aps,prb,showpacs]{revtex4-1}
\UseRawInputEncoding
\usepackage{graphicx}
\usepackage[CJKbookmarks,dvipdfm,colorlinks,linkcolor=blue,citecolor=blue]{hyperref}

\begin{document}

\title{Piezoelectric quantum spin Hall insulator VCClBr monolayer with pure out-of-plane piezoelectric response}

\author{San-Dong Guo$^{1}$,   Wen-Qi Mu$^{1}$, Hao-Tian Guo$^{1}$,  Yu-Ling Tao$^{1}$ and  Bang-Gui Liu$^{2,3}$}
\affiliation{$^1$School of Electronic Engineering, Xi'an University of Posts and Telecommunications, Xi'an 710121, China}

\affiliation{$^2$ Beijing National Laboratory for Condensed Matter Physics, Institute of Physics, Chinese Academy of Sciences, Beijing 100190, People's Republic of China}
\affiliation{$^3$School of Physical Sciences, University of Chinese Academy of Sciences, Beijing 100190, People's Republic of China}
\begin{abstract}
The combination of piezoelectricity with  nontrivial topological insulating phase in two-dimensional (2D) systems, namely piezoelectric quantum spin Hall insulator (PQSHI), is intriguing  for exploring novel
topological states toward the development of high-speed and
dissipationless electronic devices.
In this work, we predict a PQSHI Janus monolayer VCClBr constructed from $\mathrm{VCCl_2}$, which is  dynamically, mechanically  and thermally stable.
In the absence of  spin orbital coupling (SOC), VCClBr is a narrow gap semiconductor with gap value of 57 meV, which is different from Dirac semimetal $\mathrm{VCCl_2}$.
The gap of VCClBr is due to built-in electric field caused by asymmetrical upper and lower atomic layers, which is further confirmed by external-electric-field induced gap in $\mathrm{VCCl_2}$.
When including SOC, the gap of VCClBr is improved to 76 meV, which is  larger than
the thermal energy of  room temperature (25 meV).  The VCClBr is  a 2D topological insulator (TI), which is confirmed by $Z_2$ topological invariant and nontrivial one-dimensional edge states. It is proved that the  nontrivial  topological properties of VCClBr are robust against strain (biaxial  and uniaxial cases) and external electric field. Due to broken horizontal mirror symmetry, only out-of-plane piezoelectric response can be observed, when   biaxial  or uniaxial in-plane strain is applied.  The predicted  piezoelectric strain coefficients $d_{31}$ and $d_{32}$ are  -0.425 pm/V  and -0.219 pm/V, which  are higher than or  compared with ones of many 2D materials. Finally, another two Janus monolayer VCFBr and VCFCl (dynamically unstable) are constructed, and they are still PQSHIs. Moreover, their $d_{31}$ and $d_{32}$ are higher than ones of VCClBr, and the $d_{31}$  (absolute value) of VCFBr is larger than one. According to out-of-plane piezoelectric coefficients  of  VCXY (X$\neq$Y=F, Cl and  Br), $\mathrm{CrX_{1.5}Y_{1.5}}$ (X=F, Cl and  Br; Y=I) and  NiXY (X$\neq$Y=Cl, Br and I), it is concluded that the size of  out-of-plane piezoelectric coefficient has a positive relation with electronegativity difference of X and Y atoms.
Our works enrich the diversity of  Janus 2D materials, and open a new avenue in searching for
PQSHI with large out-of-plane piezoelectric response, which  provides a potential platform  in nanoelectronics.

\end{abstract}
\keywords{Piezoelectricity, Topological insulator,  Janus structure~~~~~~~~~~Email:sandongyuwang@163.com}

\maketitle

\section{Introduction}
The discovery of  2D  materials has
spurred a surge of experimental and theoretical interests in
understanding their physical and chemical properties.
The piezoelectricity of 2D materials has been a research hotspot in the ever-increasing energy conversion area\cite{q4,q4-1}, which requires a 2D system to be non-centrosymmetrical.  The 2H-$\mathrm{MoS_2}$ is a typical 2D piezoelectric material, which has been confirmed experimentally\cite{q5}.
In theory, piezoelectric properties of many 2D materials have been reported\cite{q11,q7,q10,q9,q13,q17,q16}. Strong in-plane piezoelectric response is common, while large out-of-plane piezoelectric strain coefficient is rare.  Searching for  2D materials  with large out-of-plane piezoelectric response has become an important quest  due to its compatibility with bottom/top gate technologies.  Another interesting feature of 2D systems is nontrivial topology, and TI of them is also called as quantum spin Hall insulator (QSHI) with spin-momentum-locked conducting
edge states. The first QSHI is predicted in graphene\cite{t3}, and then the  HgTe/CdTe and InAs/GaSb quantum wells are  experimentally confirmed as QSHIs \cite{t4,t5}.
Theoretically, many 2D systems have been proposed as QSHIs by first-principle calculations\cite{t6,t7,qt3,t8,t9,t10}.
Searching for QSHI with large gap (larger than 25 meV) is very important for room temperature device applications.
\begin{figure*}
  \includegraphics[width=12.0cm]{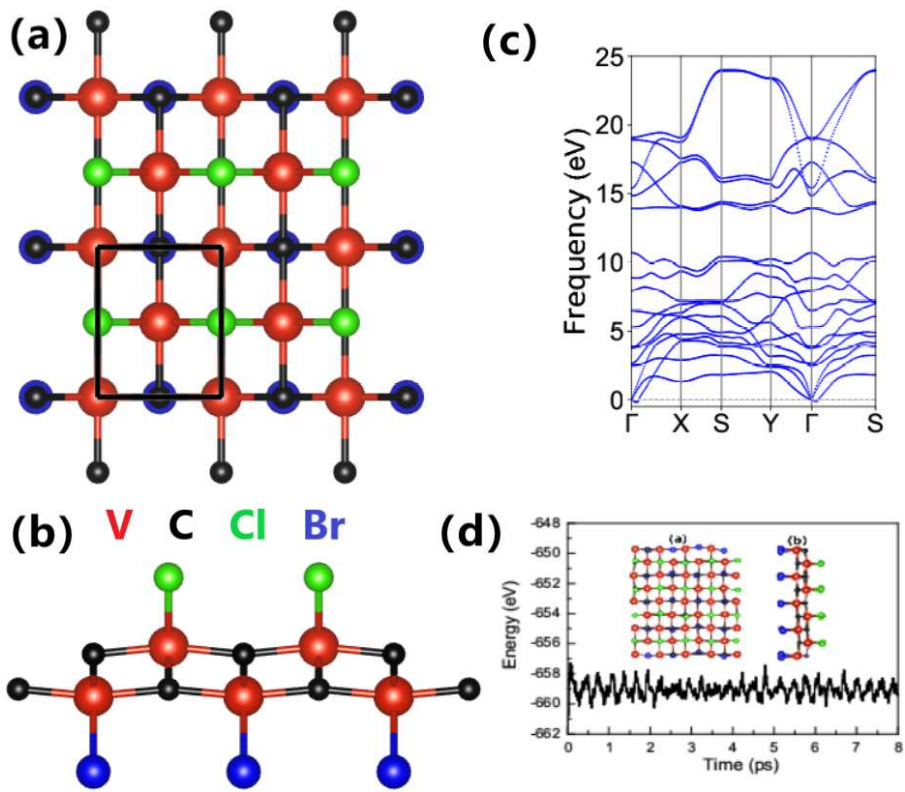}
  \caption{(Color online)For Janus monolayer VCClBr, the (a) top view and (b) side view of crystal structure, and   the   primitive cell  is marked  by  black  frames;  (c):The phonon band dispersions  by using GGA; (d):The  total energy fluctuations as a function of simulation time at 300 K, and insets show the
 final structures (top view (a) and side view (b)) of VCClBr after 8 ps at 300 K. }\label{t0}
\end{figure*}

Multifunctional 2D materials  are of particular interest due to providing  a potential platform for multi-functional electronic devices.
Compared to individual  piezoelectricity or  nontrivial band topology  in a single 2D material, their combination  can provide a unique opportunity
for intriguing physics and practical device applications. To search PQSHI,  there are two ways :(1) seeking for nontrivial band topology from 2D piezoelectric materials; (2) finding or constructing non-centrosymmetrical QSHI, which may be a executable method, especially for sandwich structure.
It is well known that a  sandwich structure can be used as a parent to construct Janus structure, which lacks central symmetry, giving rise to  piezoelectric effect\cite{gsd1,gsd2,gsd4}.  The typical Janus MoSSe monolayer has been  experimentally synthesized based on $\mathrm{MoS_2}$ with symmetrical upper and lower atomic layers\cite{e1,e2}.
However, the  nontrivial topology may be broken by building Janus structure from a QSHI\cite{gsd4,gsd4-1}.
The  Janus  $\mathrm{SrAlGaSe_4}$ and MoSSe as derivatives of QSHI $\mathrm{SrGa_2Se_4}$ and 1T'-$\mathrm{MoS_2}$ lose nontrivial topology. However, strain   can  make them possess topological properties of their parents\cite{gsd4,gsd4-1}.

Recently, based on  VOCl as a prototype,  20 2D monolayers VXY (X = B, C, N, O and F; Y = F, Cl, Br and I) are predicted by means of first-principles calculations\cite{w-1}.
They  possess  emerging topological, magnetic and dielectric properties, which depend electron occupation, ligand
electronegativity and crystal field strength. Calculated results demonstrate that the VCY
systems exhibit nontrivial topological properties\cite{w-1}.  Their crystal structure possesses  central symmetry,  which leads to disappeared piezoelectricity.  However, this structure can be regarded as a sandwich structure. These provide a possibility to induce  PQSHI by constructing Janus structure  based on VCY monolayer as parent.

In this work, we build a Janus crystal structure VCClBr, which can be attained by  replacing the  top Cl layer in  $\mathrm{VCCl_2}$  monolayer with Br atoms.
Calculated results show that  VCClBr indeed is a PQSHI with  dynamical, mechanical  and thermal  stabilities. It is proved that nontrivial band topology  of VCClBr is
robust against strain and  external electric field. The calculated out-of-plane piezoelectric response is strong compared with ones of many 2D materials\cite{q7,q9}, like Janus transition metal dichalcogenide (TMD) monolayers. Finally, by analysing  out-of-plane piezoelectric coefficients of  VCXY (X$\neq$Y=F, Cl and  Br), $\mathrm{CrX_{1.5}Y_{1.5}}$ (X=F, Cl and  Br; Y=I)\cite{w-2} and  NiXY (X$\neq$Y=Cl, Br and I)\cite{w-3}, the strong  out-of-plane piezoelectric response can be attained  by  constructing Janus monolayer with large electronegativity difference of X and Y atoms.
Our works highlight  multifunctionalities of Janus VCClBr, which is
promising for applications in the next-generation topotronic and  piezoelectric devices.

The rest of the paper is organized as follows. In the next
section, we shall give our computational details. In  the next few sections,  the  crystal structure and stability, electronic structures and  piezoelectric properties of Janus  monolayer  VCClBr are shown.  Finally,  our discussion and conclusions are given.

\begin{table}
\centering \caption{For monolayer VCClBr, VCFBr and VCFCl, the lattice constants $a$ and $b$ ($\mathrm{{\AA}}$), GGA+SOC gaps $E^g_{SOC}$ (meV), band gap category $BGC$ [direct bandgap (D) and indirect bandgap (I)],  topological
invariant $Z_2$ and dynamical stability.}\label{tab}
  \begin{tabular*}{0.48\textwidth}{@{\extracolsep{\fill}}ccccccc}
  \hline\hline
Name&$a$ &$b$&$E^g_{SOC}$&$BGC$&$Z_2$&Dynamics\\\hline
VCClBr & 3.257& 3.959&76&D&1&Y\\\hline
VCFBr & 3.140&3.949&45&I&1&N\\\hline
VCFCl&3.092&3.949 & 20&D&1&N\\\hline\hline
\end{tabular*}
\end{table}

\section{Computational detail}
We perform density functional theory (DFT) calculations\cite{1} using the Vienna Ab initio
Simulation Package (VASP), with projected augmented wave method\cite{pv1,pv2,pv3}.
  The total energy  convergence criterion and cutoff energy
for plane-wave expansion are set
to $10^{-8}$ eV and 500 eV, respectively.  The exchange-correlation potential is adopted within generalized gradient approximation (GGA) of Perdew, Burke and  Ernzerhof\cite{pbe}. Furthermore, Heyd-Scuseria-Ernzerhof screened hybrid functional
(HSE06)\cite{hse} is also employed to confirm nontrivial topology. The SOC is included  to investigate electronic structures and topological properties of VCClBr. To sample the Brillouin zone (BZ),  a Monkhorst-Pack mesh of 16$\times$13$\times$1  is adopted for geometry optimization, and  the residual force on each atom  is required to be  less than 0.0001 $\mathrm{eV.{\AA}^{-1}}$.
 To eliminate
spurious interactions with its periodic images, a vacuum region
of more than 18 $\mathrm{{\AA}}$ is adopted in the direction perpendicular
to the VCClBr monolayer.

The phonon dispersion
spectrums  are attained  by finite displacement method  with a supercell
of 5$\times$5$\times$1, as implemented in Phonopy code\cite{pv5}.  To calculate the second order interatomic force constants (IFCs), a 2$\times$2$\times$1 k-mesh is employed with kinetic energy cutoff of 500 eV. The elastic stiffness tensor $C_{ij}$ and  piezoelectric stress coefficients $e_{ij}$ are attained by using  strain-stress relationship (SSR) and density functional perturbation theory (DFPT) method\cite{pv6}, respectively.
The 2D elastic coefficients $C^{2D}_{ij}$
 and   piezoelectric stress coefficients $e^{2D}_{ij}$
have been renormalized by the the length of unit cell along $z$ direction ($L_z$):  $C^{2D}_{ij}$=$L_z$$C^{3D}_{ij}$ and $e^{2D}_{ij}$=$L_z$$e^{3D}_{ij}$.
A 16$\times$13$\times$1 Monkhorst-Pack  mesh  is adopted to calculate  $C_{ij}$ and $e_{ij}$ by using  GGA.
WannierTools code\cite{w1} is used to calculate  surface state and $Z_2$ topological invariant, based on the tight-binding Hamiltonians constructed from maximally localized Wannier functions  by employing $d$-orbitals of V atoms and  $p$-orbitals of C, Cl and Br atoms, as  implemented in Wannier90 code\cite{w2}

\begin{figure}
  \includegraphics[width=8cm]{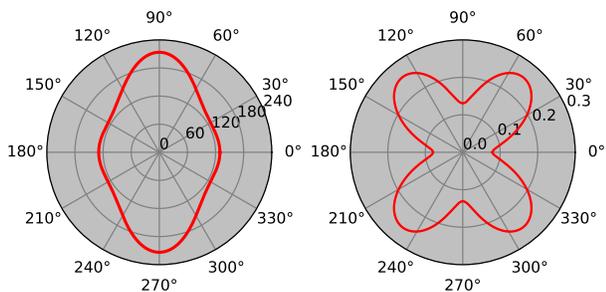}
\caption{(Color online) Angular dependence of the Young's modulus $C_{2D}(\theta)$ (Left) and Poisson's
ratio $\nu_{2D}(\theta)$ (Right) of  monolayer VCClBr.}\label{cv}
\end{figure}

\begin{figure}
  \includegraphics[width=8cm]{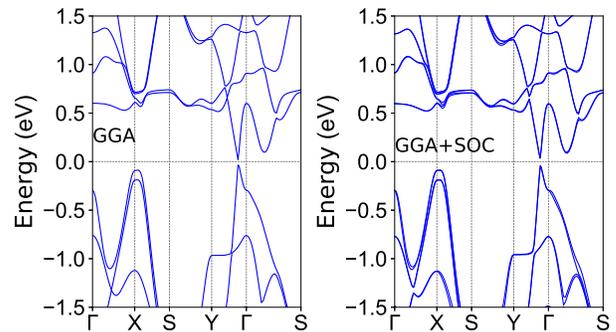}
\caption{(Color online)Top:The energy band structures  of  VCClBr  using GGA  and GGA+SOC.}\label{band}
\end{figure}

\begin{figure*}
   \includegraphics[width=12cm]{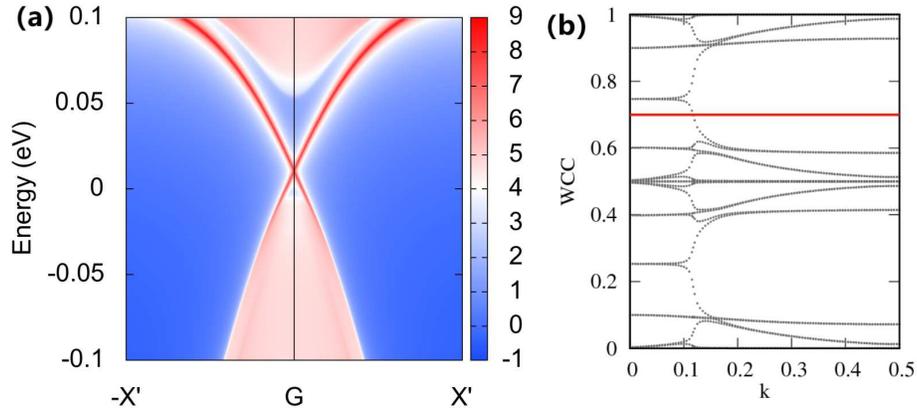}
\caption{(Color online) The nontrivial topological edge states (Left) and evolution of WCC (Right) of VCClBr monolayer. }\label{band-s}
\end{figure*}

\begin{figure*}
  \includegraphics[width=12cm]{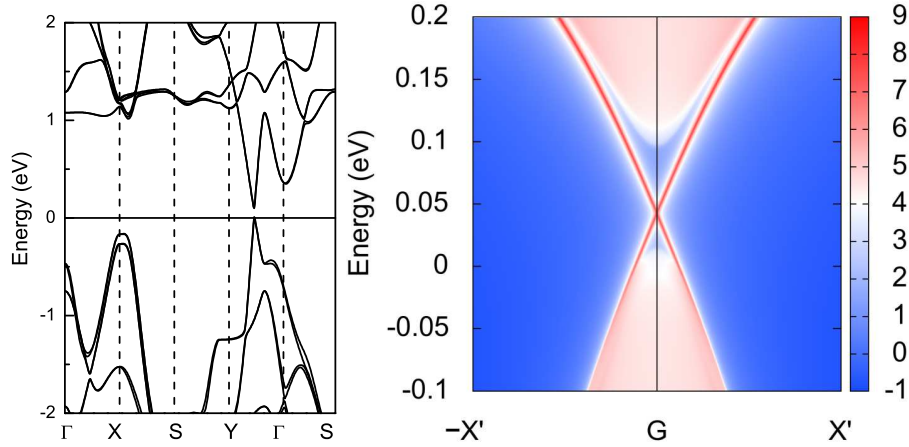}
\caption{(Color online)The energy band structures (Left)  of  VCClBr  using HSE06 along with the corresponding nontrivial topological edge states (Right).}\label{band-hse}
\end{figure*}

\begin{figure}
  \includegraphics[width=8cm]{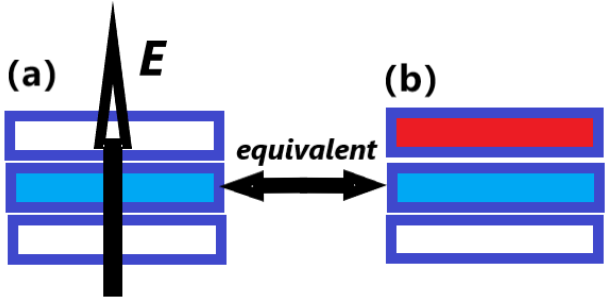}
\caption{(Color online)The schematic layered crystal structures of  $\mathrm{VCCl_2}$ (a) and VCClBr (b). Applying  an external electric field $E$ on $\mathrm{VCCl_2}$  can be used to simulate Janus structure VCClBr. }\label{e}
\end{figure}

\begin{figure}
   \includegraphics[width=7.0cm]{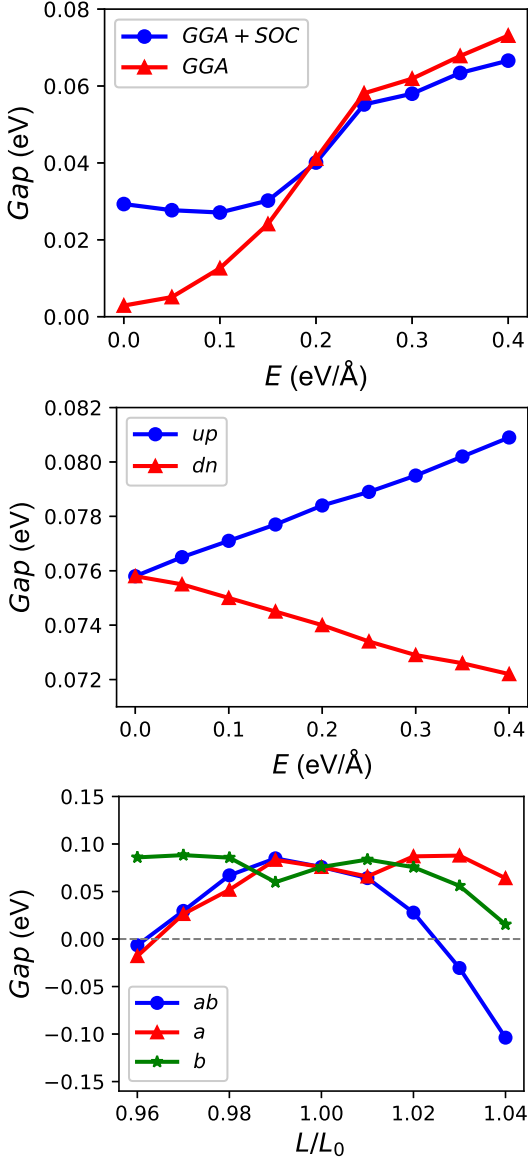}
  \caption{(Color online) Top plane: the energy band gaps of  $\mathrm{VCCl_2}$ monolayer  as a function of applied electric field  by using both GGA and GGA+SOC; Middle plane: the energy band gaps of  VCClBr monolayer  as a function of applied electric field along +$z$ (up) and -$z$ (dn) directions  by using  GGA+SOC; Bottom plane: the energy band gaps of  VCClBr monolayer  as a function of   applied strain, including biaxial (ab) and uniaxial cases (a and b).  }\label{gap}
\end{figure}
\begin{figure}
  \includegraphics[width=8cm]{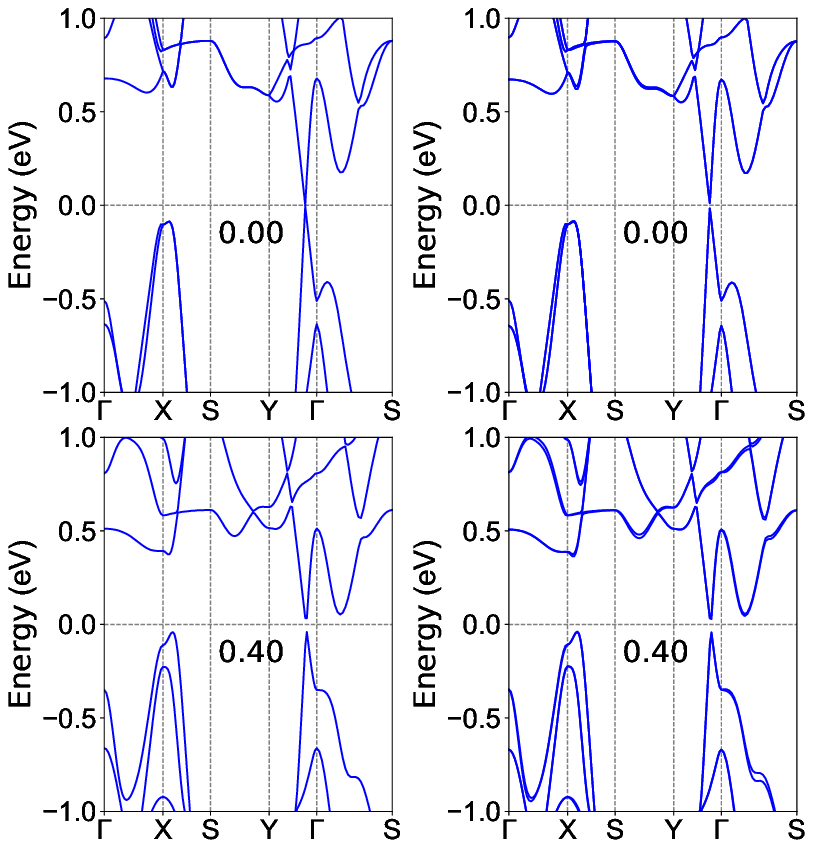}
\caption{(Color online) The  energy bands of   $\mathrm{VCCl_2}$ monolayer  under 0.00 eV/$\mathrm{{\AA}}$ and 0.04
eV/$\mathrm{{\AA}}$ electric field  using GGA (Left)  and GGA+SOC (Right).}\label{cl-e-band}
\end{figure}

\section{Crystal structure and stability}
As shown in \autoref{t0} (a) and (b), the middle
VC layer of VCClBr is sandwiched by Cl and Br layers, which can be considered as a Janus structure.  The VCClBr  can be constructed by  replacing  top Cl atom layer  in  $\mathrm{VCCl_2}$  monolayer with Br atoms. Its each V atom is 6-
fold coordinated with four C , one Cl and one  Br atoms, giving rise to a strongly distorted octahedron unit.
The symmetry of VCClBr (No.25) is lower than that of  $\mathrm{VCCl_2}$  (No.59) due to
the lack of  horizontal mirror symmetry. Namely, the $\mathrm{VCCl_2}$ has centrosymmetry with disappeared piezoelectricity, while the VCClBr lacks  inversion symmetry, possessing piezoelectricity. The optimized lattice
constant of VCClBr is  a=3.257 $\mathrm{{\AA}}$ and  b=3.959 $\mathrm{{\AA}}$. Due to different atomic sizes and electronegativities of Cl and Br atoms,  inequivalent  bond lengths can be observed, for example $d_{Cl-V}$=2.449 $\mathrm{{\AA}}$ and  $d_{Br-V}$=2.575 $\mathrm{{\AA}}$, which  can induce an electrostatic potential gradient,  giving rise to a built-in electric field.

To validate its dynamic stability, the phonon dispersions  of VCClBr  are calculated, which is plotted in \autoref{t0} (c).
No imaginary vibrational frequency with three acoustic
and fifteen optical phonon branches is observed,  suggesting that VCClBr is
dynamically stable. In addition,  by performing ab-initio molecular dynamics (AIMD) simulations,  the thermal stability  of VCClBr with a supercell of  4$\times$4$\times$1 for more than 8000 fs  at 300 K is examined. The total energy fluctuations as a function of simulation time along with the final
configurations are plotted in \autoref{t0} (d). Calculated results show that VCClBr undergoes  no structural
reconstruction with  small energy
 fluctuates,   indicating its thermal stability.

 To further check
the mechanical stability of VCClBr, its elastic constants are calculated.
By using Voigt notation, the elastic tensor with $mm2$ point-group symmetry  can be written as:
\begin{equation}\label{pe1-4}
   C=\left(
    \begin{array}{ccc}
      C_{11} & C_{12} & 0 \\
     C_{12} & C_{22} &0 \\
      0 & 0 & C_{66} \\
    \end{array}
  \right)
\end{equation}
The calculated  $C_{11}$=130.67 $\mathrm{Nm^{-1}}$, $C_{22}$=215.81 $\mathrm{Nm^{-1}}$, $C_{12}$=17.08 $\mathrm{Nm^{-1}}$  and  $C_{66}$=53.45 $\mathrm{Nm^{-1}}$.
The calculated elastic constants satisfy the  Born  criteria of mechanical stability:
 \begin{equation}\label{pe1-4-0}
  C_{11}>0,~~C_{22}>0,~~ C_{66}>0,~~C_{11}-C_{12}>0
\end{equation}
 Therefore, VCClBr is  mechanically stable.

Since the $C_{11}$ and  $C_{22}$ of  VCClBr are very different, the  anisotropy
of its Young's moduli $C_{2D}(\theta)$ and
Poisson's ratios $\nu_{2D}(\theta)$ should be very distinct. The direction-dependent $C_{2D}(\theta)$ and $\nu_{2D}(\theta)$  can be attained by  the following two formulas\cite{ela,ela1}:
 \begin{equation}\label{pe1-4-1}
  C_{2D}(\theta)=\frac{C_{11}C_{22}-C_{12}^2}{C_{11}m^4+C_{22}n^4+(B-2C_{12})m^2n^2}
\end{equation}
 \begin{equation}\label{pe1-4-2}
  \nu_{2D}(\theta)=\frac{(C_{11}+C_{22}-B)m^2n^2-C_{12}(m^4+n^4)}{C_{11}m^4+C_{22}n^4+(B-2C_{12})m^2n^2}
\end{equation}
where the $\theta$ is  the angle of direction with $x$/$y$ direction defined as as $0^{\circ}$/$90^{\circ}$, $m=sin(\theta)$, $n=cos(\theta)$ and $B=(C_{11}C_{22}-C_{12}^2)/C_{66}$.
 As plotted in \autoref{cv}, the  $C_{2D}(\theta)$ and
 $\nu_{2D}(\theta)$ of  VCClBr  show very strong anisotropy.
The $C_{2D}$ is 129.32/213.58  $\mathrm{Nm^{-1}}$ along $x$/$y$ direction, and 0.079/0.131 for  $\nu_{2D}$.

\begin{figure*}
   \includegraphics[width=16cm]{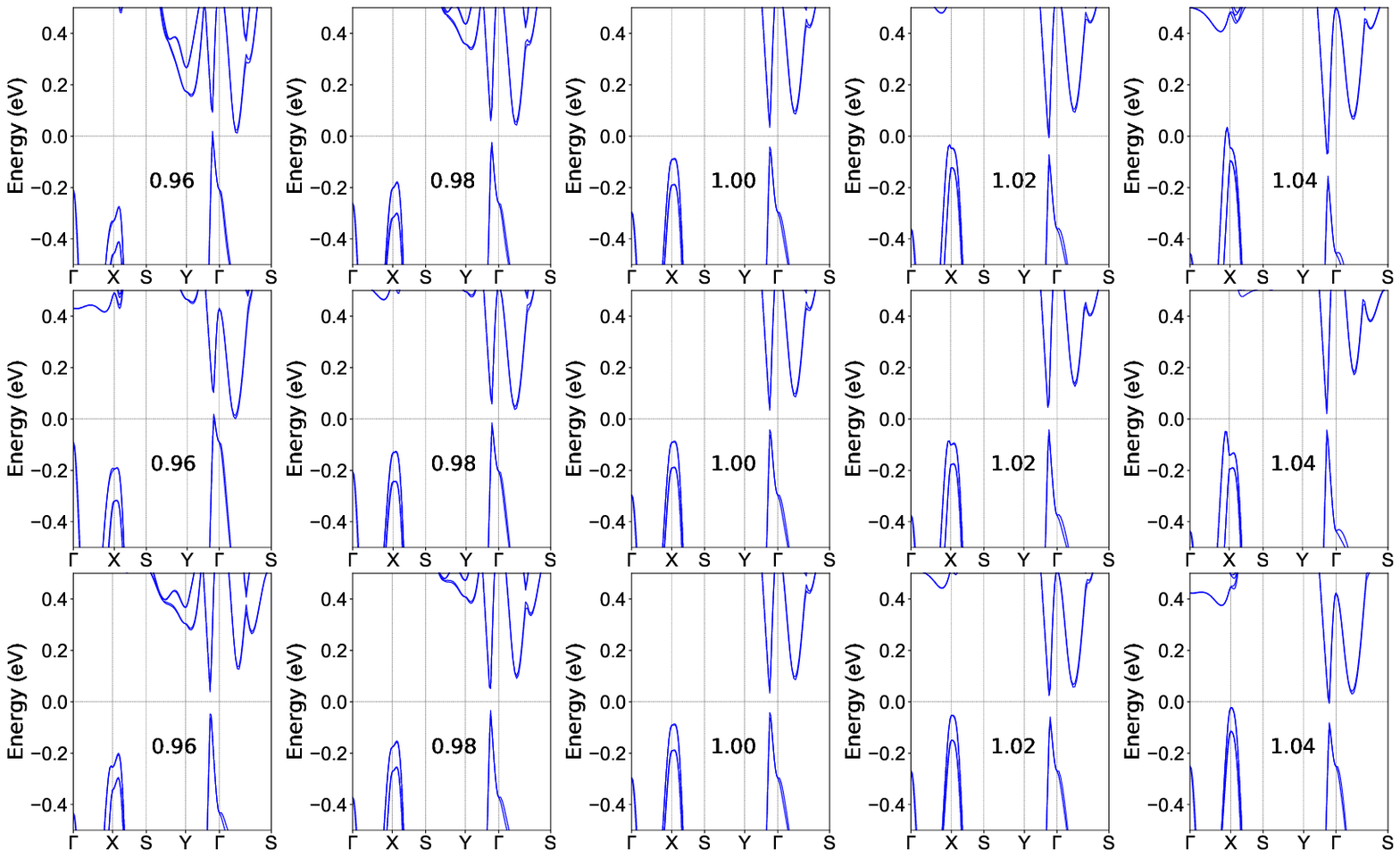}
\caption{(Color online) The energy band structures of VCClBr with SOC  at representative applied strain, including biaxial (Top panel), uniaxial-$a$ (Middle panel) and  uniaxial-$b$ (Bottom panel) cases. }\label{band-strain}
\end{figure*}

\begin{figure}
  \includegraphics[width=8cm]{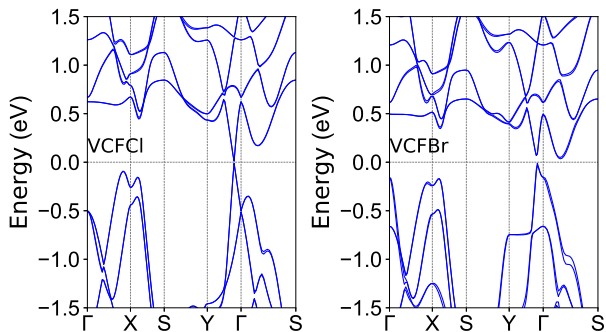}
\caption{(Color online) The energy band structures  of VCFCl and VCFBr monolayers by using  GGA+SOC. }\label{band-1}
\end{figure}

\begin{figure*}
   \includegraphics[width=12cm]{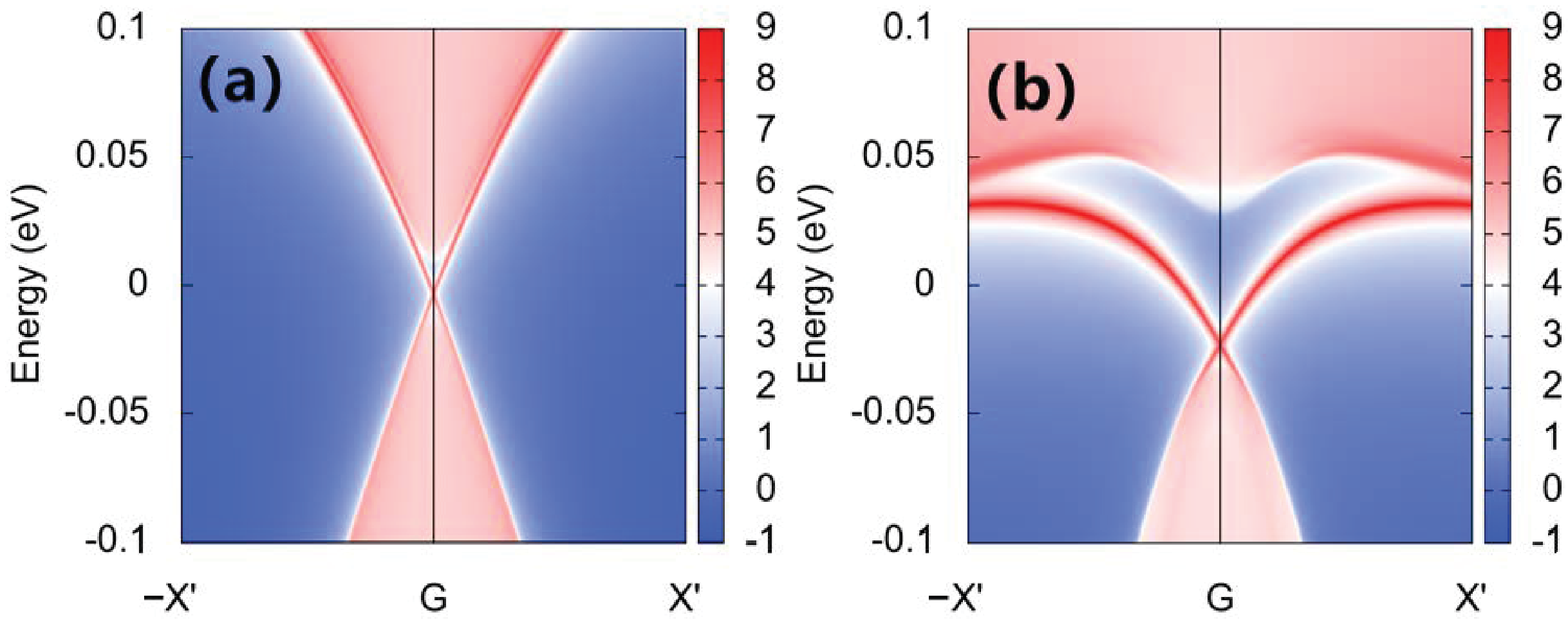}
\caption{(Color online)  The nontrivial topological edge states of VCFCl (a) and VCFBr (b).}\label{band-s-1}
\end{figure*}

\section{Electronic structures}
The energy band structures of VCClBr by using GGA and GGA+SOC are plotted in \autoref{band}.
 In the absence of SOC, the VCClBr is a direct gap semiconductor with the gap of  57 meV, and the valence band maximum (VBM) and conduction band minimum (CBM) locate  at the one point along $\Gamma$-Y path. It clearly seen that the very strong anisotropy is observed for energy band structures along $\Gamma$-X and $\Gamma$-Y paths.
  When including SOC, the VCClBr is still  a direct gap semiconductor, and the gap is improved to 76 meV.
The previous calculated results show  that $\mathrm{VCCl_2}$ is a QSHI\cite{w-1}.  The Janus monolayer  $\mathrm{SrAlGaSe_4}$ as a derivative of centrosymmetric QSHI $\mathrm{SrGa_2Se_4}$ is a normal insulator, which can be reverted by strain\cite{gsd4}. However, Janus  monolayer  $\mathrm{CSb_{1.5}Bi_{1.5}}$ still maintains nontrivial topological properties from QSHI $\mathrm{CX_3}$ (X=Sb and Bi) with inversion symmetry\cite{gsd5}. To confirm the topological properties of VCClBr,  the $Z_2$ topological
invariant is calculated. For a material with  inversion symmetry,  $Z_2$ topological
invariant can be attained from the product of parities of all
occupied states at four time-reversal-invariant-momentum
 points in the 2D BZ. However, VCClBr is non-centrosymmetrical,  and the $Z_2$  can be verified  by the calculation of
Wannier charge center (WCC). If $Z_2$ equals 1/0, a material is  a topologically nontrivial/trivial state.
As plotted in \autoref{band-s},   an arbitrary reference line crosses  the evolution lines of WCC  an odd number of times, giving rise to $Z_2$=1, which indicates
that VCClBr is a QSHI. Furthermore, a QSHI should  exhibit non-trivial topological
edge states.  We further calculate one-dimensional edge states on (100) edge, which are plotted in \autoref{band-s}.
It is clearly seen that topological helical edge states with the appearance of the Dirac cone  connect the conduction and valence bands.
The HSE06 is also employed to check the reliability of nontrivial topological properties of VCClBr. Within HSE06+SOC, the energy band structure and one-dimensional edge states are plotted in \autoref{band-hse}. Calculated results show that the gap of VCClBr is improved to 87 meV from 76 meV with GGA+SOC, and the nontrivial topological properties are still kept with topological helical edge states.

The $\mathrm{VCCl_2}$ is a  Dirac semimetal without considering  SOC, and no opened gap is observed\cite{w-1}.
Compared to  $\mathrm{VCCl_2}$,  VCClBr lacks  horizontal mirror symmetry,  possessing a built-in electric field, which induces the opened gap.
To confirm this, we apply  an external electric field ($E$) perpendicular to the $\mathrm{VCCl_2}$, which should induce a gap.
 Since the electric field is a  vector (not a pseudovector),  the horizontal mirror symmetry will be broken by electric field.
 This is different from the magnetization as a pseudovector, which preserves the horizontal mirror symmetry.
Thus, applying  an external electric field can be used to simulate Janus structure (see \autoref{e}) in a sense.

For $\mathrm{VCCl_2}$ monolayer, the evolution of  band gap as a function of applied electric field is provided in \autoref{gap} by using GGA and GGA+SOC, and the representative energy bands    under 0.00 eV/$\mathrm{{\AA}}$ and 0.04
eV/$\mathrm{{\AA}}$ electric field are plotted in \autoref{cl-e-band}.  In the absence of SOC,  the band gap increases  monotonously from 0 meV with increasing electric field for $\mathrm{VCCl_2}$. This implies that applied electric field can induce the gap for $\mathrm{VCCl_2}$,   which is  equivalent to a built-in electric field by constructing Janus structure, like VCClBr. For example 0.04
eV/$\mathrm{{\AA}}$ electric field, the gap of $\mathrm{VCCl_2}$  is  73 meV without considering SOC, and the GGA+SOC gap is reduced to 67 meV.
It is noted that the gap with GGA+SOC is larger than one with GGA under small electric field, and then the opposite situation is observed with sequentially increasing electric field.  Finally, the edge states are calculated within considered $E$ range.  It is found that the nontrivial topological properties of $\mathrm{VCCl_2}$ are robust against applied electric field, and the representative topological helical edge states under  0.40
eV/$\mathrm{{\AA}}$ electric field are plotted in FIG.1 of electronic supplementary information (ESI).

The electric field effects on electronic and topological properties of VCClBr are also investigated.  Janus VCClBr has different atomic species on its upper and lower facets (see \autoref{e}), which means that applying $+z$ and $-z$ directional electric field are not equivalent.  This is different from $\mathrm{VCCl_2}$, which  has the same atomic species on its upper and lower facets (see \autoref{e}), giving rise to equivalent effects  caused by applying both $+z$ and $-z$ directional electric field.
 The energy band gaps of  VCClBr monolayer  as a function of applied electric field along +$z$  and -$z$  directions  by using  GGA+SOC are plotted in \autoref{gap}.
The gap increases with increasing  $+z$  directional $E$, while increasing  $-z$  directional $E$ reduces the gap.  However, the gap change is very small with $E$ from 0.00 eV/$\mathrm{{\AA}}$ to 0.40 eV/$\mathrm{{\AA}}$, and only
5.1/-3.6 meV for $+z$ and $-z$ directional $E$, which is very smaller than one (37.3 meV) of  $\mathrm{VCCl_2}$.
It is also proved that the nontrivial topological properties of VCClBr are robust against applied electric field.

Strain is an effective method to tune electronic properties of 2D systems, and further affects their topological properties\cite{h1,h2,h3}.
Here, both biaxial  and uniaxial strains are considered to investigate the robustness of nontrivial topological properties of VCClBr.
 The $L/L_0$  is used to simulate compressive/tensile strain, where $L$ and $L_0$ are the strained and  unstrained lattice constants, respectively.   The $L/L_0$$<$1 ($L/L_0$$>$1) means compressive (tensile) strain. For biaxial strain,  the $a$  and $b$ lattice parameters simultaneously change, and atomic positions are allowed to relax. For uniaxial strain,  the $a$/$b$ lattice parameter changes, and the $b$/$a$ lattice parameter and atomic positions are allowed to relax  in response to
an applied in-plane uniaxial strain.

The electronic structures  versus $L$/$L_0$ are calculated by using GGA+SOC. The variation of band gaps with biaxial, uniaxial-$a$ and  uniaxial-$b$  cases are plotted in \autoref{gap}, and the energy band structures at representative applied strain are shown in \autoref{band-strain}.
For biaxial strain, with increasing strain, the gap firstly increases, and then decreases, which is due to the transition of CBM from one point  along $\Gamma$-S to  one point along $\Gamma$-Y. It is found that both compressive and tensile strains can induce semiconductor-metal transitions, which are due to the variation of CBM and VBM.
Our calculated results show that the nontrivial topological properties are robust within considered strain range.
For  0.96 or 1.04 strain, VCClBr is a metal, but the gap along  $\Gamma$-Y path is still maintained, and topological helical edge states still exist (see FIG.2 of ESI). Partly, the  uniaxial-$a$ compressive/tensile strain is equivalent to  uniaxial-$b$ tensile/compressive one, which can be confirmed by their gaps versus $L/L_0$.
It is found that  uniaxial strain can also induce the change of CBM or VBM, giving rise to nonmonotonic dependence of gap on $L/L_0$. Semiconductor-metal transitions can also be observed at  uniaxial-$a$ 0.96 strain.  However, the gap along  $\Gamma$-Y path always exists. The topological helical edge states can always be observed with considered uniaxial strain range, which are plotted in FIG.3 and FIG.4 of ESI at 0.96 and 1.04 strains.

\begin{table*}
\centering \caption{For monolayer VCClBr, VCFBr and VCFCl, the elastic constants $C_{ij}$ ($\mathrm{Nm^{-1}}$), piezoelectric stress coefficient $e_{ij}$ along with electronic part $e_{ij_e}$ and  ionic part $e_{ij_i}$ ($10^{-10}$ C/m), and piezoelectric strain coefficient $d_{ij}$ (pm/V). }\label{tab0}
\begin{tabular*}{0.96\textwidth}{@{\extracolsep{\fill}}ccccccccccccc}
  \hline\hline
Name& $C_{11}$&	$C_{12}$&	$C_{22}$&	$C_{66}$&	$e_{31_e}$&	$e_{31_i}$&	$e_{31}$&	$e_{32_e}$&	$e_{32_i}$&	$e_{32}$&	$d_{31}$&	$d_{32}$\\\hline
VCClBr& 130.67&	17.08&	215.81&	53.45&	-0.612&	0.019&	-0.593&	-0.513&	-0.032&	-0.545&	-0.425&	-0.219\\\hline
VCFBr& 151.06&	21.42&	229.97&	54.47&	-1.957&	0.115&	-1.842&	-1.669&	-0.243&	-1.912&	-1.116&	-0.727\\\hline
VCFCl& 161.47&	23.10&	229.17&	55.06&	-1.359&	0.165&	-1.194&	-1.135&	-0.176&	-1.311&	-0.667&	-0.505\\\hline\hline
\end{tabular*}
\end{table*}

\section{Piezoelectric properties}
 For  noncentrosymmetric material, an  applied strain or stress can induce  electric
dipole moments and produce electricity, called  piezoelectric effect. The pristine  $\mathrm{VCCl_2}$ is non-piezoelectric, while VCClBr with particular Janus structure  possessess piezoelectric effect.
The third-rank piezoelectric stress tensor  $e_{ijk}$ and strain tensor $d_{ijk}$ can be used to describe piezoelectric response of a material.  The relaxed piezoelectric tensors ($e_{ijk}$ and $d_{ijk}$) are obtained as the sum of ionic and electronic contributions:
 \begin{equation}\label{pe0}
      e_{ijk}=\frac{\partial P_i}{\partial \varepsilon_{jk}}=e_{ijk}^{elc}+e_{ijk}^{ion}
 \end{equation}
and
 \begin{equation}\label{pe0-1}
   d_{ijk}=\frac{\partial P_i}{\partial \sigma_{jk}}=d_{ijk}^{elc}+d_{ijk}^{ion}
 \end{equation}
Where $P_i$, $\varepsilon_{jk}$ and $\sigma_{jk}$ are polarization vector, strain and stress, respectively, and the superscripts $elc$ and $ion$ mean electronic and ionic contributions. The  $e_{ijk}^{elc}$ and $d_{ijk}^{elc}$  ($e_{ijk}$ and $d_{ijk}$) are also called clamped-ion (relax-ion) piezoelectric coefficients.
The $e_{ijk}$ and $d_{ijk}$ can be connected  by elastic tensor $C_{mnjk}$:
 \begin{equation}\label{pe0-1-1}
    e_{ijk}=\frac{\partial P_i}{\partial \varepsilon_{jk}}=\frac{\partial P_i}{\partial \sigma_{mn}}.\frac{\partial \sigma_{mn}}{\partial\varepsilon_{jk}}=d_{imn}C_{mnjk}
 \end{equation}

For 2D materials,  these conditions ( $\varepsilon_{ij}$=$\sigma_{ij}$=0 for i=3 or j=3) are assumed, which implies that only the in-plane strain and stress are taken into account. However, the polarization can be allowed  to remain out-of-plane. We
define the in-plane directions $x_1$ and $x_2$ as the  short edge and long edge direction of  primitive cell,
and $x_3$ as perpendicular to the 2D layer.
Employing Voigt notation,  the  piezoelectric stress   and strain tensors  due to $mm2$ point group symmetry  can be reduced into :
 \begin{equation}\label{pe1-1}
 e=\left(
    \begin{array}{ccc}
     0 & 0 & 0 \\
     0 & 0 & 0 \\
      e_{31} & e_{32} & 0 \\
    \end{array}
  \right)
    \end{equation}

  \begin{equation}\label{pe1-2}
  d= \left(
    \begin{array}{ccc}
      0 & 0 & 0 \\
       0 & 0 & 0 \\
      d_{31} & d_{32} &0 \\
    \end{array}
  \right)
\end{equation}
When   a uniaxial in-plane strain is applied,  only a  vertical piezoelectric polarization  exists ($d_{31}$$\neq$0 or $d_{32}$$\neq$0).
Moreover,  with applied  biaxial in-plane strain,  the superposed
out-of-plane  polarization  will arise ($d_{31}$$\neq$0 and $d_{32}$$\neq$0).
The $e_{31}$ and  $e_{32}$  can be directly  calculated by DFPT, and the
$d_{31}$ and  $d_{32}$   can be  derived by \autoref{pe1-4}, \autoref{pe0-1-1}, \autoref{pe1-1} and \autoref{pe1-2}:
\begin{equation}\label{pe2-0}
     d_{31}=\frac{e_{31}C_{22}-e_{32}C_{12}}{C_{11}C_{22}-C_{12}^2}
\end{equation}
\begin{equation}\label{pe2-1}
     d_{32}=\frac{e_{32}C_{11}-e_{31}C_{12}}{C_{11}C_{22}-C_{12}^2}
\end{equation}

The calculated $e_{31}$ ($e_{32}$) is -0.593$\times$$10^{-10}$ C/m (-0.545$\times$$10^{-10}$ C/m)  with ionic part 0.019$\times$$10^{-10}$ C/m (-0.032$\times$$10^{-10}$ C/m)  and electronic part -0.612$\times$$10^{-10}$ C/m (-0.513$\times$$10^{-10}$ C/m).
The electronic and ionic polarizations  of $e_{31}$ have  opposite signs,  while they have the same signs for $e_{32}$. However, the electronic contribution
 dominates the  piezoelectricity of both $e_{31}$ and $e_{32}$.  Based on \autoref{pe2-0} and \autoref{pe2-1}, the $d_{31}$  and $d_{32}$  can be attained  from previous calculated $C_{ij}$ and $e_{ij}$. The calculated  $d_{31}$ ($d_{32}$)  is  -0.425 pm/V (-0.219 pm/V),  which are  higher than or  compared with ones of  other 2D materials\cite{q7,q9}.   The $d_{31}$ is almost twice as large as $d_{32}$, which is mainly  due to very different $C_{11}$  and $C_{22}$.  Thus,  VCClBr  is a potential PQSHI.

\begin{figure}
   \includegraphics[width=8cm]{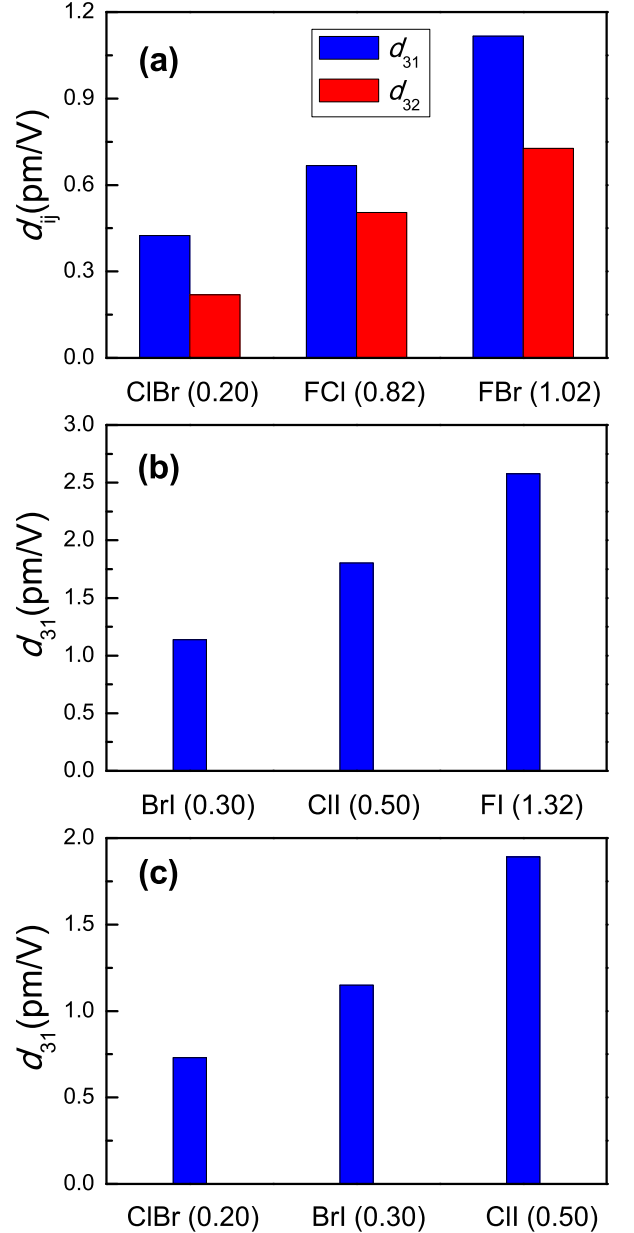}
\caption{(Color online)  The $d_{31}$  and  $d_{32}$ of (a): VCXY (X$\neq$Y=F, Cl and  Br), (b):$\mathrm{CrX_{1.5}Y_{1.5}}$ (X=F, Cl and  Br; Y=I) and (c): NiXY (X$\neq$Y=Cl, Br and I) . The electronegativity difference of X and Y atoms are given in parentheses.}\label{d}
\end{figure}
\section{Discussions and Conclusion}
 The $\mathrm{VCY_2}$ (Y=F, Cl, Br and I) monolayers have been predicted to be topological Dirac semimetals without SOC\cite{w-1}. However, for $\mathrm{VCI_2}$,  the Fermi level deviates from Dirac point. Thus, based on $\mathrm{VCY_2}$ (Y=F, Cl and Br) monolayers, we construct another two Janus monolayer VCFBr and VCFCl of VCXY (X$\neq$Y=F, Cl and  Br).
 The optimized lattice constants $a$ and $b$ are listed in \autoref{tab}. It is found that  $a$ increases with the size of  X and Y anions
(from VCFCl to VCFBr to VCClBr), while $b$ is less affected, which is because the
X and Y anions are directly bonded to V atoms along  $x$ direction. The energy band structures of VCFBr and VCFCl within SOC are plotted in \autoref{band-1}.
Calculated results show that VCFCl and VCClBr are direct gap semiconductors, while VCFBr is an indirect gap semiconductor. The CBM of VCFBr is at one point along $\Gamma$-S path, while the VBM locates at one point along $\Gamma$-Y path. For  VCFCl and VCClBr, both CBM and VBM locate at one point along $\Gamma$-Y path.
As summarized in \autoref{tab}, the gap increases with the size of X and Y anions
(from VCFCl to VCFBr to VCClBr). It is clearly seen that topological helical edge states connect the valence and conduction bands of VCFBr and VCFCl (see \autoref{band-s-1}), confirming their nontrivial topological properties.

 For monolayer VCClBr, VCFBr and VCFCl, the calculated elastic constants $C_{ij}$, piezoelectric stress coefficient $e_{ij}$ along with electronic   and  ionic parts, and piezoelectric strain coefficient $d_{ij}$ are listed in \autoref{tab0}. For these monolayers, calculated elastic constants  satisfy   mechanical stability criteria, confirming their mechanical stabilities.
For all these monolayers, the electronic and ionic contributions  of $e_{31}$   have  opposite signs,  while they have the same signs for $e_{32}$, and the electronic part dominates  $e_{31}$  and  $e_{32}$. For three monolayers, the $d_{31}$ is larger than $d_{32}$.  The $d_{31}$ and $d_{32}$ of VCFBr are the largest  among these monolayers, which may be due to very different electronegativities of F and Br atoms compared with F and Cl or Cl and Br atoms. The $d_{31}$ and $d_{32}$ (absolute value) of VCXY (X$\neq$Y=F, Cl and  Br) along with the electronegativity difference of X and Y atoms are plotted in \autoref{d}.
  Similar phenomenon  can be found in monolayer $\mathrm{CrX_{1.5}Y_{1.5}}$ (X=F, Cl and Br; Y=I) and NiXY (X$\neq$Y=Cl, Br and I)\cite{w-2,w-3}, and  their $d_{31}$  along with the electronegativity difference of X and Y atoms are also plotted in \autoref{d}. It is clearly seen that the large  electronegativity difference of X and Y atoms is related with large out-of-plane piezoelectric response. It is noted that the $d_{31}$ of VCFBr is larger than 1.0 pm/V, which is rare for out-of-plane piezoelectric response of 2D systems.   Finally, the phonon dispersions of  VCFBr and VCFCl are calculated, which are plotted in FIG.5 of ESI. Unfortunately, the two monolayers are dynamically unstable due to existing imaginary frequencies. However, these comparisons provide some thoughts and scientific basis for  searching large out-of-plane piezoelectric response in Janus 2D materials.

In summary, we have demonstrated coexistence of piezoelectricity and robust nontrivial band topology in Janus VCClBr monolayer with dynamical, mechanical and thermal  stabilities based on the reliable DFT calculations. The  nontrivial band topology in Janus VCClBr monolayer is confirmed by topological helical edge states, which is
robust against strain and external electric field.
 Only out-of-plane piezoelectric response exists due to broken horizontal mirror symmetry. The predicted   out-of-plane $d_{31}$ and $d_{32}$  are  higher than  or comparable with ones of  many 2D known materials, which is highly desirable for ultrathin piezoelectric devices. It is found that the elastic, electronic and piezoelectric properties of  VCClBr monolayer show very strong anisotropy due to the Cl and Br anions  directly bonded to V atoms along the $x$ direction.
Our predicted VCClBr provides a possible platform  to future development
of multifunctional piezoelectronics, and these findings open
new idea to realize  PQSHI with large out-of-plane piezoelectric response.

\begin{acknowledgments}
This work is supported by Natural Science Basis Research Plan in Shaanxi Province of China  (2021JM-456). We are grateful to the Advanced Analysis and Computation Center of China University of Mining and Technology (CUMT) for the award of CPU hours and WIEN2k/VASP software to accomplish this work.
\end{acknowledgments}

\end{document}